\documentclass[aps,prd,preprint,preprintnumbers,unsortedaddress,superscriptaddress,showpacs,nofootinbib]{revtex4-1}

 \usepackage{graphicx,color}
\usepackage{relsize}
\usepackage{slashed}
\usepackage{color}
\usepackage{ulem}
\usepackage{tabu}
\usepackage{bm}
\usepackage{rotating}

\usepackage{amsmath}
\usepackage{amssymb}
\usepackage{amsthm}
\usepackage{mathrsfs}
\usepackage{graphicx}
\usepackage{epstopdf}
\usepackage{fancyhdr}
\usepackage{array}
\usepackage[all]{xy}
\usepackage{eufrak}
\usepackage{euscript}
\usepackage{enumerate}
\usepackage{slashed}
\usepackage{hyperref}
\usepackage{subfigure} % NEEDED IN ORDER TO USE SUBFIG
\usepackage{epstopdf} % COMPILE EPS (AND PDF) FIGURES USING PDFLATEX!!!

\hypersetup{pdftex,colorlinks=true,linkcolor=blue,citecolor=blue,menucolor=black,urlcolor=blue,filecolor=blue}

\begin{document}

\title{$\chi_{c1}$ decays into a pseudoscalar meson and a vector-vector molecule}

\author{Natsumi Ikeno}
\email{ikeno@tottori-u.ac.jp}
\affiliation{Department of Physics, Guangxi Normal University, Guilin 541004, China}
\affiliation{
Department of Life and Environmental Agricultural Sciences,
Tottori University, Tottori 680-8551, Japan}
\affiliation{Departamento de F\'isica Te\'orica and IFIC, Centro Mixto Universidad de Valencia-CSIC, Institutos de Investigaci\'on de Paterna, Aptdo. 22085, 46071 Valencia, Spain}

\author{Jorgivan M. Dias}
\email{jdias@if.usp.br}
\affiliation{Department of Physics, Guangxi Normal University, Guilin 541004, China}
\affiliation{Instituto de F\'{i}sica, Universidade de S\~{a}o Paulo, Rua do Mat\~{a}o, 1371, Butant\~{a}, CEP 05508-090, S\~{a}o Paulo, S\~{a}o Paulo, Brazil}

\author{Wei-Hong Liang}
\email{liangwh@gxnu.edu.cn}
\affiliation{Department of Physics, Guangxi Normal University, Guilin 541004, China}

\author{Eulogio Oset}
\email{eulogio.oset@ific.uv.es}
\affiliation{Department of Physics, Guangxi Normal University, Guilin 541004, China}
\affiliation{Departamento de F\'isica Te\'orica and IFIC, Centro Mixto Universidad de Valencia-CSIC, Institutos de Investigaci\'on de Paterna, Aptdo. 22085, 46071 Valencia, Spain}

\date{\today}

\begin{abstract}
We evaluate ratios of the $\chi_{c1}$ decay rates to $\eta$ ($\eta', K^-$) and one of the $f_0(1370)$, $f_0(1710)$, $f_2(1270)$, $f_2'(1525)$, $K^{*}_2(1430)$ resonances, which in the local hidden gauge approach are dynamically generated from the vector-vector interaction. With the simple assumption that the $\chi_{c1}$ is a singlet of SU(3), and the input from the study of these resonances as vector-vector molecular states, we describe the experimental ratio $\mathcal{B}(\chi_{c1} \rightarrow \eta f_2(1270))/ \mathcal{B}(\chi_{c1} \rightarrow \eta' f_2'(1525))$ and make predictions for six more ratios that can be tested in future experiments. 
\end{abstract}

%\pacs{Valid PACS appear here}
% PACS, the Physics and Astronomy Classification Scheme.
% Valid PACS numbers may be entered using the \verb+\pacs{#1} command.

% \keywords{Meson-baryon interaction; $\Omega_b$ states; Molecular state}

\maketitle
%\tableofcontents

%%%%%%%%%%%%%%%%%%%%%%%%%%%%%%%%%%
\section{Introduction}
\label{sec:intro}
The topic of hadronic resonances is capturing the attention of hadron physics and gradually the field is piling up more evidence that some hadronic states stand for a molecular interpretation in terms of more elementary hadrons. Recent reviews on the topic can be seen in~\cite{guo,hanhartxyz}.

A good laboratory to see such states in the light quark sector is the decay of charmonium states into three lighter mesons. Indeed the $c\bar{c}$ states can be considered as an SU(3) singlet, in the same way that an $s\bar{s}$ state is an isospin singlet.
With the SU(3) matrices for the mesons one can construct SU(3) invariants, which provide trios of mesons with a certain weight. If a given resonance is formed from the interaction of pairs of mesons, the picture to produce them is to produce first these mesons and then let them interact through final state interaction where the resonances will be formed.
This picture has been thoroughly used.
Indeed, in~\cite{ollerpsi} the $J/\psi \rightarrow \phi(\omega) f_0(980) (f_0(500))$ decays were studied producing $\phi(\omega)$ and $K\bar{K}$, $\pi\pi$, $\eta\eta$ pairs and allowing the $K\bar{K}$, $\pi\pi$, $\eta\eta$ pairs to interact to produce the $f_0(500)$, $f_0(980)$ states, which are well described within the chiral unitary approach as dynamically generated from the interactions of these pairs of mesons~\cite{npa,Kaiser,markushin,juan}.
A different formalism of the same problem, although equivalent, is used in~\cite{palomar}, which is then followed in~\cite{lahde}.
The same $J/\psi$ decay into three mesons is considered in~\cite{liangsakai} with $J/\psi \rightarrow VPP$, but letting the $VP$ pairs interact to form an axial-vector meson. A mixture of the SU(3) trace $ \langle VPP \rangle$  and $ \langle V \rangle \langle PP \rangle$ was used and it was shown to correspond to the same structures developed in \cite{ollerpsi} and \cite{palomar} and there it was used to describe successfully the BESIII data on the $J/\psi \rightarrow \eta(\eta') h_1 (1380)$ decay~\cite{besh1}. The $\langle VPP \rangle$ term was shown to be the one corresponding to the dominant one in~\cite{ollerpsi,palomar}.

The dominance of the three meson matrix trace seems to be quite general in these processes since in the study of the $\chi_{c1} \rightarrow \eta \pi^+ \pi^-$ reaction leading to the formation of the $f_0(500)$, $f_0(980)$ and $a_0(980)$ resonances~\cite{Kornicer}, the $\langle PPP \rangle$ production structure was found largely dominant and the final state interaction of the different $PP$ pairs led to the production of these scalar resonances~\cite{liangchi1,debastiani}.

We have described processes that proceed as a first step from the $VPP$ and the $PPP$ production. The processes proceeding from $VVV$ production were studied in~\cite{alberdai} in the $J/\psi \rightarrow \phi(\omega) f_2(1270)$ and related reactions. In this case one of the vectors $\phi(\omega)$ acts as a spectator and the remaining $VV$ pair produces the $f_2(1275)$, $f'_2(1525)$ resonances, which according to~\cite{raquel,gengvec} are produced mostly from the $\rho \rho$ and $K^* \bar{K^*}$ interaction, respectively\footnote{The ability of the methods used in refs.~\cite{raquel,gengvec} to obtain the tensor states has been questioned in \cite{gulmez} and \cite{gulguo}. In \cite{raquerep} and \cite{gengrep}, a thorough discussion of these works has been done, showing that the methods proposed cannot be extrapolated to the energies where the tensor states appear. At the same time an improved method is proposed that provides couplings of the resonances to the coupled channels practically identical to those of \cite{raquel,gengvec} which we use here. }. A fresh look to this latter problem from the new perspective of $\langle VVV \rangle$ and $\langle VV \rangle \langle V \rangle$ production vertices, including more experimental data, has been given in~\cite{raquelrong}, where a detailed discussion on molecular states is made in the Introduction.

As to processes proceeding with $VVP$ production we can have two cases. One of them corresponds to having a $V$ as spectator and $VP$ interacting producing an axial vector meson. An example of this is the $\chi_{c1}$ decay to $\phi h_1 (1380)$~\cite{beschicj}, which has been described successfully along the lines described above with the $\langle VVP \rangle$ structure~\cite{shengjuan}. There is only one more case to be studied which corresponds to the same structure but now the $P$ will be a spectator and the $VV$ will interact to form the vector-vector molecular states of~\cite{raquel,gengvec}, $f_2(1270)$, $f_2'(1525)$, $f_0(1370)$, $f_0(1710)$, $K^*_2(1430)$. We undertake this work here and compare the results with the few data available on $\eta$ ($\eta', \bar{K}$) production together with one of these resonances.

%%%%%%%%%%%%%%%%%%%%%%%%%%%%%%%%%%
\section{Formalism}
\label{sec:form}

We study reactions that can be compared with present results in the PDG~\cite{pdg}. This includes $\chi_{c1}$ decays with $\eta, \eta'$ production and some of the $f_0$, $f_2$ resonances, but we shall make predictions for related decays which are likely to be measured in the near future. As advanced in the Introduction, we introduce the $\langle VVP \rangle$ and $\langle VV \rangle \langle P \rangle$ structures in the primary step, where $V$, $P$ are the SU(3) vector and pseudoscalar matrices (corresponding to $q\bar{q}$) given by

\begin{equation}
V = \left(
           \begin{array}{ccc}
             \frac{\rho^0}{\sqrt{2}} + \frac{\omega}{\sqrt{2}}  & \rho^+ & K^{*+} \\
             \rho^- & -\frac{\rho^0}{\sqrt{2}} + \frac{\omega}{\sqrt{2}}  & K^{*0} \\
            K^{*-} & \bar{K}^{*0} & \phi \\
           \end{array}
         \right),
\label{eq:Vmatrix}
\end{equation}

\begin{equation}
P = \left(
           \begin{array}{ccc}
             \frac{\pi^0}{\sqrt{2}} + \frac{\eta}{\sqrt{3}} + \frac{\eta'}{\sqrt{6}} & \pi^+ & K^+ \\
             \pi^- & -\frac{\pi^0}{\sqrt{2}} + \frac{\eta}{\sqrt{3}} + \frac{\eta'}{\sqrt{6}} & K^0 \\
            K^- & \bar{K}^0 & -\frac{1}{\sqrt{3}}\eta + \sqrt{\frac{2}{3}}\eta' \\
           \end{array}
         \right),
\label{eq:phimatrix}
\end{equation}
where we have assumed the $\eta$, $\eta'$ mixing according to~\cite{bramon}. The algebra for $\langle VVP \rangle$ is trivial, and isolating the terms containing $\eta$, $\eta'$ or $K^-$ we find the production structures:

 \begin{eqnarray}
& & 1) \ \eta: \  \frac{\eta}{\sqrt{3}} \left\{ \rho^0 \rho^0 + \rho^+ \rho^- + \rho^- \rho^+ + \omega \omega - \phi \phi \right\};\\ 
& & 2)\  \eta' : \ \frac{\eta'}{\sqrt{6}} \left\{ \rho^0 \rho^0 + \rho^+ \rho^- + \rho^- \rho^+ + \omega \omega + 3 K^{*+} K^{*-} + 3 K^{*0}\bar{K}^{*0} +2 \phi \phi \right\}; \\
& & 3) \ K^- : \ K^-  \left\{   \left( \frac{\rho^0}{\sqrt{2}} + \frac{\omega}{\sqrt{2}} \right)  K^{*+} + \rho^+ K^{*0} + K^{*+} \phi   \right\}.
\end{eqnarray}

In order to connect with the isospin formalism of~\cite{raquel,gengvec} we write the isospin states with the unitary normalization with our phase convention, $(-\rho^+, \rho^0, \rho^-)$, ($K^{*+}, K^{*0}$), ($\bar{K}^{*0}, -K^{*-}$), as
\begin{eqnarray}
& & |\rho \rho, I=0 \rangle  = -\frac{1}{\sqrt{6}} |\rho^0 \rho^0 + \rho^+ \rho^- + \rho^- \rho^+ \rangle, \\
& & | \omega \omega, I=0 \rangle  =  \frac{1}{\sqrt{2}} | \omega \omega \rangle, \\
& & | \phi \phi, I=0 \rangle  =  \frac{1}{\sqrt{2}} | \phi \phi \rangle, \\
& & | \rho K, I=\frac{1}{2}, I_3 = \frac{1}{2} \rangle  =  -\sqrt{\frac{2}{3}} \rho^+ K^{*0} - \sqrt{\frac{1}{3}} \rho^0 K^{*+}.
\end{eqnarray}

Taking into account the symmetry factor $n!$ for production of $n$ identical particles we find the weights for primary production of the different components as

\begin{eqnarray}
& &  h_{\rho \rho}^{(\eta)} = -\sqrt{\frac{1}{2}} ;\hspace{5mm} h_{\omega \omega}^{(\eta)} = \sqrt{\frac{2}{3}} ;\hspace{5mm} h_{\phi \phi}^{(\eta)} = -\sqrt{\frac{2}{3}}; \hspace{5mm} h_{K^* \bar{K}^*}^{(\eta)} =0;   \nonumber\\
& & h_{\rho \rho}^{(\eta')} = -\frac{1}{2} ;\hspace{5mm} h_{\omega \omega}^{(\eta')} = \sqrt{\frac{1}{3}} ;\hspace{5mm} h_{\phi \phi}^{(\eta)'} =  \frac{2}{\sqrt{3}};\ h_{K^* \bar{K}^*}^{(\eta')} =- \frac{\sqrt{3}}{2};   \nonumber\\
& & h_{\rho K^*}^{(K^-)} = -\sqrt{\frac{3}{2}} ;\hspace{5mm} h_{\omega K^*}^{(K^-)} = \sqrt{\frac{1}{2}} ;\hspace{5mm} h_{\phi K^*}^{(K^-)} = 1 . 
\label{eq:p7.2}
\end{eqnarray}

In addition we will also consider the $\langle VV \rangle \langle P \rangle$ structure, which could mix with the expected dominant  $\langle VVP \rangle$ one with some small admixture.
Now we have
\begin{eqnarray}
 \langle VV \rangle \langle P \rangle = \left(  \frac{\eta}{\sqrt{3}} + \frac{4\eta'}{\sqrt{6}}\right)
(\rho^0 \rho^0 + \rho^+ \rho^- + \rho^- \rho^+ + \omega \omega + \phi \phi + 2K^{*+} K^{*-} + 2 K^{*0} \bar{K}^{*0}).
\nonumber\\
\end{eqnarray}

We see that there is no contribution for $K^-$ production with this term and we find the weights for $\eta$, $\eta'$ production as
\begin{eqnarray}
& &  h'^{(\eta)}_{\rho \rho} = h^{(\eta)}_{\rho \rho} ; \hspace{5mm} h'^{(\eta)}_{\omega \omega} = h^{(\eta)}_{\omega \omega}; \hspace{5mm} h'^{(\eta)}_{\phi \phi} = -h^{(\eta)}_{\phi \phi}; \hspace{5mm} h'^{(\eta)}_{K^* \bar{K}^*} = -\sqrt{\frac{2}{3}}; \nonumber\\ 
& &  h'^{(\eta')}_{\rho \rho} = 4 h^{(\eta')}_{\rho \rho}; \hspace{5mm} h'^{(\eta')}_{\omega \omega} = 4 h^{(\eta')}_{\omega \omega}; \hspace{5mm} h'^{(\eta')}_{\phi \phi} = 2h^{(\eta')}_{\phi \phi}; \hspace{5mm} h'^{(\eta')}_{K^* \bar{K}^*} = \frac{8}{3} h^{(\eta')}_{K^* \bar{K}^*}. 
\end{eqnarray}

\begin{figure}[h!]\centering
  \includegraphics[scale=0.5]{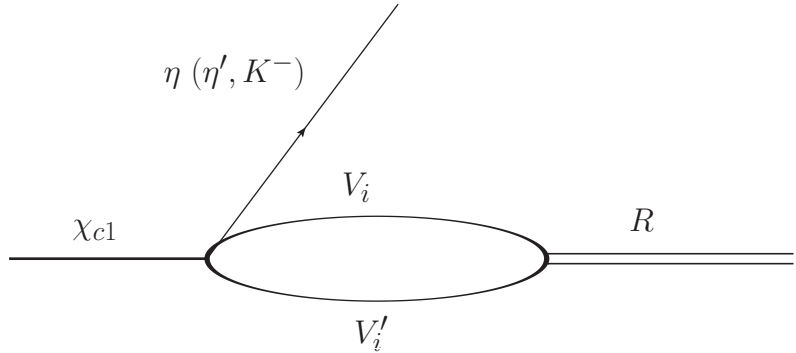}
\caption{Mechanism for $\chi_{c1}$ decay into $\eta (\eta', K^-)$ and the resonance $R$, via intermediate $V_i$, $V'_i$ production and coupling of these states to the resonance. 
}
\label{fig:1}
\end{figure}

We will assume a structure
\begin{equation}
 \langle VVP \rangle + \beta \langle VV \rangle \langle P \rangle\, ,
\label{eq:p7.1}
\end{equation}
and, hence, in order to take into account the second term of Eq.~(\ref{eq:p7.1}) one simply has to substitute 
\begin{equation}
 h^{(\eta)}_i \rightarrow h^{(\eta)}_i + \beta h'^{(\eta)}_i ;\  \hspace{1cm}
 h^{(\eta')}_i \rightarrow h^{(\eta')}_i + \beta h'^{(\eta')}_i.
\label{eq:p7.3}
\end{equation}
Note that there is no contribution to $K^-$ production from the $\langle VV \rangle \langle P \rangle$ structure.
The final state interaction is taken into account as described diagrammatically in Fig.~\ref{fig:1}.

Analytically the mechanism of Fig.~\ref{fig:1} leads to the $\chi_{c1}$ decay amplitude
\begin{equation}
 t'^{(\eta)}_{R} = \sum_{i} h^{(\eta)}_i G_i (M_R) g_{R,i}\, ,
\end{equation}
and the same for $\eta'$, $K^-$ production, where $h^{(\eta)}_i$ are the weights given in Eq.~(\ref{eq:p7.2}), $G_i(M_{R})$ is the loop function for the pair of intermediate vectors, and $g_{R, i}$ are the couplings of the resonance $R$ to the different $V_i V'_i$ intermediate channels. All these magnitudes are evaluated in~\cite{raquel,gengvec} and for completeness they are given here in Table~\ref{table:1}. When considering the combination of Eq.~(\ref{eq:p7.1}) we shall make the replacement of Eq.~(\ref{eq:p7.3}).

We still have to make some considerations concerning the spin of the $VV$ states and angular momentum conservation of the production vertex. The $\chi_{c1}$ has 
$J^{PC}=1^{++}$, the $\eta(\eta^{\prime})$ $0^{-+}$ and the vector-vector states that we consider are $0^{++}$, $2^{++}$. $C-$parity is conserved but in order to conserve parity 
and total angular momentum we need a $P-$wave. Since we require $S-$wave for the propagation of the two vectors, the momentum involved must be the one of the pseudoscalar. 
Thus
\begin{equation}\label{Tamp}
t^{(\eta)}_R=t^{\prime\,(\eta)}_R\ \vec{\epsilon}_{\chi_{c1}}\cdot \vec{p}_{\eta}\, ,
\end{equation}
and similarly for the other cases, where 
\begin{equation}\label{pcm}
|\vec{p}_{\eta}| =\frac{\lambda^{1/2}(M^2_{\chi_{c1}},M^2_{\eta},M^2_R)}{2M_{\chi_{c1}}}\,.
\end{equation}
In addition we should take into account that the $V_iV_i^{\prime}$ pair is produced with spin $0$ or $2$ for which the proper amplitudes are provided in 
\cite{raquel,gengvec}. However, their consideration only introduces spin factors after summing over spins, and since we only compare rates 
for $VV^{\prime}$ states with the same spin, $f_0(1370)$ with $f_0(1710)$, $f_2(1270)$ with $f^{\prime}_2(1525)$ etc., it is 
unnecessary to consider these extra factors. Note that we have studied the flavor dependence of the production vertex to relate the different components, but the 
production rate could be different for the production of $VV^{\prime}$ pairs with different spin. The $P-$wave structure of Eq.~\eqref{Tamp} is important since it introduces a $\vec{p\,\,}^2_{\eta}\,(\vec{p\,\,}^2_{\eta^{\prime}},\, \vec{p\,\,}^2_{K^-})$ factor in $|t_R|^2$ and this factor is quite different for 
$\eta$ or $\eta^{\prime}$ for instance.

The decay width is given by
\begin{equation}
\Gamma^{(\eta)}_R=\frac{1}{8\pi}\frac{1}{M^2_{\chi_{c1}}}\mathcal{C}\, |t^{\prime\,(\eta)}_R|^2\, {p}^2_{\eta}\,\, p_{\eta}\, ,
\label{Gamma}
\end{equation}
with $p_{\eta}$ given by Eq.~\eqref{pcm}, and similarly for the other cases. In Eq.~\eqref{Gamma} $\mathcal{C}$ is a global normalization constant, which 
cancels in the ratios that we discuss.

\begin{sidewaystable}
\caption{ Table of $g$, $G$ values of the different channels and different resonances, like in \cite{alberdai} (including the $f_0(1370)$ $f_0(1710)$) and errors).}
\vspace{0.3cm}
\centering
\begin{small}
\renewcommand{\arraystretch}{0.5}
\begin{tabular}{l c| c ccccccc}
\hline\hline
&&&&&&&&&\\
 &&  &$\rho\rho$&$K^* \bar{K}^*$&$\omega\omega$&$\phi\phi$&$\rho\bar{K}^*$&$\omega\bar{K}^{*\,0}$&$\phi\bar{K}^{*\,0}$\\
[0.1ex]
\hline
% Entering 1st row
 & &$g_{i} (\textrm{MeV})$ &10551 & $4771$ & $-503$ & $-771$&0&0&0   \\[1ex]
 & & error of $g_{i} (\%) $
&  $4$ & $3$ & $22$ & 22&0&0&0  \\[1.ex]
\raisebox{1.5ex}{}  &\raisebox{1.5ex}{$f_{2}(1270)$} &$G_{i} (\times 10^{-3})$
&  $-4.74$ & $-3.00$ & $-4.97$ & 0.475&0&0&0  \\[1.ex]
& &error of $G_{i} (\%)$
&  $10$ & $29$ & $42$ & 220&0&0&0  \\[1.ex]

 \hline
% Entering 2nd row
& &$g_{i} (\textrm{MeV})$ & $-2611$ & $9692$ & $-2707$ & $-4611$&0&0&0   \\[1ex]
 & & error of $g_{i} (\%) $
&  $12$ & $6$ & $2$ & 2&0&0&0  \\[1.ex]

\raisebox{1.5ex}{} & \raisebox{1.5ex}{$f^\prime_{2}(1525)$}& $G_{i} (\times 10^{-3})$
&$-8.67$ & $-4.98$ & $-9.63$ & $-0.710$&0&0&0 \\[1ex]
& &error of $G_{i} (\%)$
&  $6$ & $17$ & $19$ & 141&0&0&0  \\[1.ex]

\hline
% Entering 3rd row
& &$g_{i} (\textrm{MeV})$ & 0 & $0$ & 0 & 0&~~~10613~~~&~~~2273~~~&~~~$-2906$~~~   \\[1ex]
 & & error of $g_{i} (\%) $
&  $0$ & $0$ & $0$ & 0&3&5&5  \\[1.ex]
\raisebox{1.5ex}{} & \raisebox{1.5ex}{${\bar{K}}_{2}^{*\,0}(1430)$}& $G_{i} (\times 10^{-3})$
&0& $0$ & 0 & 0&~~~$-6.41$~~~&~~~$-5.94$~~~&~~~$-2.70 $ \\[1ex]
& &error of $G_{i} (\%)$
&  $0$ & $0$ & $0$ & 0&12&19&43  \\[1.ex]

\hline
% Entering 4rd row
& &$g_{i} (\textrm{MeV})$ & $(7913.62, -1048.34)$ & $(1209.79, -414.83)$ & $(-39.29, 30.62)$ & $(12.10, 23.68)$& 0 &0 & 0   \\[1ex]
 & & error of $g_{i} (\%) $
&  $4$ & $3$ & $22$ & 22&0&0&0  \\[1.ex]
\raisebox{1.5ex}{}  &\raisebox{1.5ex}{$f_{0}(1370)$} &$G_{i} (\times 10^{-3})$
&  $(-7.69, 1.72)$ & $(-4.13, 0.26)$ & $(-8.97, 0.87)$ & $(-0.63, 0.14)$& 0&0&0  \\[1.ex]
& &error of $G_{i} (\%)$
&  $10$ & $29$ & $42$ & 220&0&0&0  \\[1.ex]

\hline
% Entering 5rd row
& &$g_{i} (\textrm{MeV})$ & $(-1029.91, 1086.85)$ & $( 7126.63, 94.28) $ &  $(-1763.75, 108.81)$ &$(-2493.76, 204.50)$ & 0 & 0& 0   \\[1ex]
 & & error of $g_{i} (\%) $
&  $12$ & $6$ & $2$ & 2&0&0&0  \\[1.ex]
\raisebox{1.5ex}{} & \raisebox{1.5ex}{$f_{0}(1710)$}& $G_{i} (\times 10^{-3})$
&$(-9.70, 6.18)$ & $(-7.68, 0.58)$ & $(-10.85, 8.19)$ & $(-2.16, 0.13)$&0&0&0 \\[1ex]
& &error of $G_{i} (\%)$
&  $6$ & $17$ & $19$ & 141&0&0&0  \\[1.ex]
\hline
\end{tabular}
\end{small}
\label{table:1}
\end{sidewaystable}

%%%%%%%%%%%%%%%%%%%%%%%%%%%%%%%%%%
\section{Results}
\label{sec:res}

In the PDG we find the following branching ratios 
\begin{eqnarray}
& & \mathcal{B}(\chi_{c1} \rightarrow \eta f_2(1270))  = (6.7\pm 1.1)\times 10^{-4}; \\
& & \mathcal{B}(\chi_{c1} \rightarrow \eta^{\prime} f_0(1710))  = (7^{~+7}_{~-5})\times 10^{-5}; \\
& & \mathcal{B}(\chi_{c1} \rightarrow \eta^{\prime} f^{\prime}_2(1525))  = (9\pm 6)\times 10^{-5}\,.
\end{eqnarray}

We can only test the ratio
\begin{equation}\label{R1exp} 
R_1=\frac{\mathcal{B}(\chi_{c1} \rightarrow \eta f_2(1270))}{\mathcal{B}(\chi_{c1} \rightarrow \eta^{\prime} f^{\prime}_2(1525))}=[3.7-26]\,\,\textrm{(centroid at 7.4)}\,.
\end{equation}
Unfortunately, the errors in the measurements are very large, but even then we can test our formalism, which 
has no free parameter for this ratio if we take first $\beta=0$. For this evaluation it is also important to consider 
the theoretical errors. We take the same errors in $g_i,\, G_i$ discussed in \cite{alberdai} and shown in Table~\ref{table:1}. To 
calculate the errors in $R_1$ we generate random numbers for $g_i,\,G_i$ within the restricted range and evaluate 
$R_1$ each time. After several runs we evaluate the average value of $R_1$ and its dispersion. We find
\begin{equation}\label{R1th}
R^{th}_1(\beta=0)=[1.1\pm 0.3]\, ,
\end{equation}
which is still lower than Eq.~\eqref{R1exp}, indicating that some admixture of the $\langle VV\rangle \langle P \rangle$ structure is 
needed. 

Before we proceed to include the contribution of the new term it is interesting to make the following observation. In \cite{gengvec} 
it was found (see Table~\ref{table:1}) that the $f_0(1370)$ and $f_2(1270)$ couple mostly to $\rho\rho$ but very weakly to $K^*\bar{K}^*$, while 
$f_0(1710)$ and $f^{\prime}_2(1525)$ couple mostly to $K^*\bar{K}^*$. If we look at the weights $h^{(\eta)}$ in Eq.~\eqref{eq:p7.2} we see 
that $h^{(\eta)}_{K^*\bar{K}^*}=0$. The $K^*\bar{K}^*$ pair is not primary produced with the $\eta$ and hence the $f_2(1710)$ and 
$f^{\prime}_2(1525)$ resonance would not be produced. This is not the case for $\eta^{\prime}$ since in Eq.~\eqref{eq:p7.2} we find that 
$h^{(\eta^{\prime})}_{K^*\bar{K}^*}=\frac{-\sqrt{3}}{2}$ and the $f_0(1710)$ and $f^{\prime}_2(1525)$ can be produced, and are indeed observed. 
The fact that in the PDG we do not find branching ratios for $\eta f_0(1710)$, $\eta f^{\prime}_2(1525)$ would then find a natural 
explanation in the dominance of the $\langle VVP \rangle$ structure. Note that if we consider the $\langle VV \rangle\langle P \rangle$ 
structure, the coefficient $h^{\prime\,(\eta)}_{K^*\bar{K}^*}$ is no longer zero. The possibly very small $\eta f_0(1710)$, 
$\eta f^{\prime}_2(1525)$ rates would point to a small admixture of the $\langle VV \rangle\langle P \rangle$ term. 

We can improve the ratio 
of Eq.~\eqref{R1th} by taking a value of 
\begin{equation}\label{beta029}
\beta=-0.28\, ,
\end{equation}
and we find 
\begin{equation}
R^{th}_1(\beta=-0.28)=7.1 \, .
\label{R1th_beta028}
\end{equation}
It is easy to make one estimate of this value of $\beta$ based on the dominance of the $\rho\rho$ channel in the $f_2(1270)$ and 
the $K^*\bar{K}^*$ channel in the $f^{\prime}_2(1525)$. By using the values $h^{(\eta)}_{\rho\rho}$, $h^{\prime\,(\eta)}_{\rho\rho}$ 
on one side and $h^{(\eta^{\prime})}_{K^*\bar{K}^*}$, $h^{\prime\, (\eta^{\prime})}_{K^*\bar{K}^*}$ we find the extra factor 
for the ratio $R^{th}_1$ of Eq.~\eqref{R1th},
\begin{equation}\label{beta}
\Big( \frac{1+\beta}{1+\frac{8}{3}\beta}\Big)^2\, ,
\end{equation}
which for $\beta=-0.267$ renders $R^{th}_1$ to the value of $7.1$ in agreement with the results of Eq.~\eqref{R1th_beta028}.

With the caveat of large uncertainties in the predictions, given the wide range of Eq.~\eqref{R1exp}, we can make predictions for other ratios 
using the value of $\beta$ in the Eq.~\eqref{beta029} and also the value $\beta=0$ in parenthesis
\begin{eqnarray}
& & R_2=\frac{\mathcal{B}[\chi_{c1} \rightarrow \eta f^{\prime}_2(1525)]}{\mathcal{B}[\chi_{c1} \rightarrow \eta f_2(1270)]}=0.16\,\,(3.8\times 10^{-3});\nonumber\\
% =0.16\,\,(3.7\times 10^{-3});\nonumber\\
& & R_3=\frac{\mathcal{B}[\chi_{c1} \rightarrow \eta f_0(1710)]}{\mathcal{B}[\chi_{c1} \rightarrow \eta f_0(1370)]}=0.060\,\,(2.3\times 10^{-2});\nonumber\\
& & R_4=\frac{\mathcal{B}[\chi_{c1} \rightarrow \eta^{\prime} f_0(1370)]}{\mathcal{B}[\chi_{c1} \rightarrow \eta^{\prime} f_0(1710)]}=0.060\,\, (0.54);\nonumber\\
& & R_5=\frac{\mathcal{B}[\chi_{c1} \rightarrow \eta^{\prime} f_2(1270)]}{\mathcal{B}[\chi_{c1} \rightarrow \eta^{\prime} f^{\prime}_2(1525)]}=4.6\times 10^{-4}\,\, (0.81);\nonumber\\
& & R_6=\frac{\mathcal{B}[\chi_{c1} \rightarrow K^- K^{*+}_2(1430)]}{\mathcal{B}[\chi_{c1} \rightarrow \eta f_2(1270)]}=10.3\,\, (4.2);\nonumber\\
& & R_7=\frac{\mathcal{B}[\chi_{c1} \rightarrow \eta f_0(1370)]}{\mathcal{B}[\chi_{c1} \rightarrow \eta^{\prime} f_0(1710)]}=10.9\,\, (1.1)\, ,
\end{eqnarray}
and we associate an error of about $30\%$ to all these ratios for a given value of $\beta$.

\begin{figure}[h!]\centering
  \includegraphics[scale=0.9]{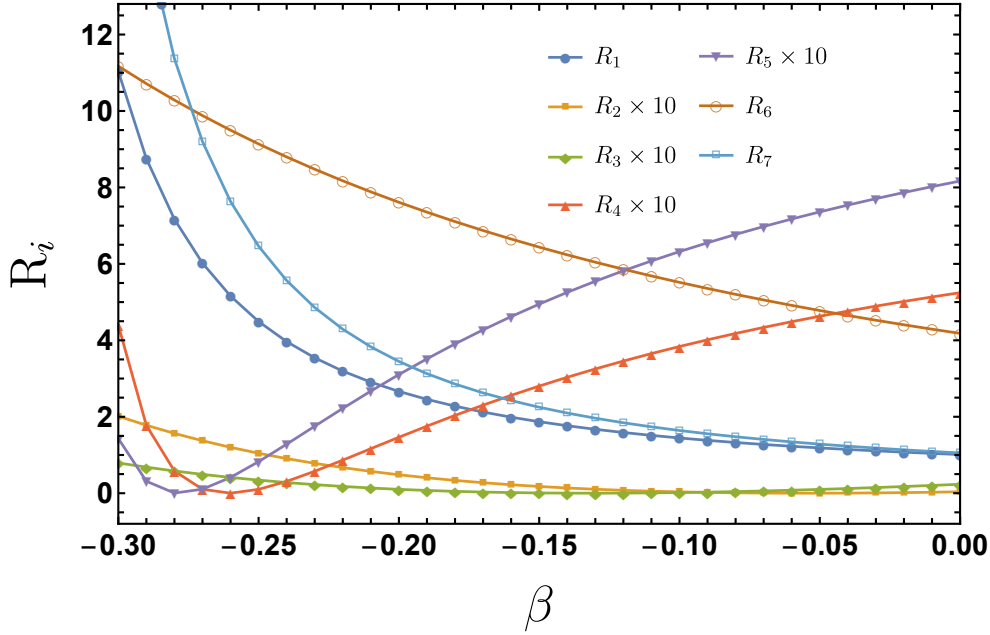}%{RX10versusBeta.eps} %{RXBeta.eps}
\caption{
Ratio $R_i$ as a function of $\beta$. 
}
\label{fig:2}
\end{figure}

We can see that $R_1$ is very sensitive to the value of $\beta$, but some of the ratios $R_2 \sim R_7$ are not so sensitive. 
The experimental situation should improve in the future and one can make the predictions more accurate. To facilitate the comparison of our predictions with future measurements, we present in Fig.~\ref{fig:2} the results of the different ratios as a function of $\beta$.

The fact that the parameter $\beta$ is negative makes the results more sensitive to the value of $\beta$, in particular for values around $-0.3$.
We also find that the behavior with $\beta$ is different for different ratios, some increase with $\beta$ and others decrease. It would be most convenient to have some other ratios measured to pin down the value of $\beta$ with some precision and make our predictions more constrained.
Only then shall we be able to appreciate the predictive power of the theory. We hope that the experimental situation is improved in the coming years and the present work should be a motivation to carry out these measurements.

\section{Conclusions}
\label{sec:conc}

We have studied reactions of $\chi_{c1}$ decay going to $\eta$ ($\eta^{\prime}$, $K^-$) and one of the resonances $f_0(1370)$, $f_2(1270)$, 
$f_0(1710)$, $f^\prime_2(1525)$ and $K^*_2(1430)$. For this we have assumed that these resonances are dynamically generated from 
the vector-vector interaction which was studied in detail in \cite{raquel,gengvec}. Then the mechanism of production proceeds via 
a primary production of a $VVP$ structure with $P=\eta,\,\eta^{\prime},\,K^-$ and the $VV$ interact later to produce the given 
resonance. Based on the simple assumption that a $c\bar{c}$ state is a singlet of $SU(3)$, in the same way as an $s\bar{s}$ state 
is a singlet of isospin, we have constructed invariants with the $SU(3)$ matrices for vectors $V$ and pseudoscalars $P$. The 
study of previous related processes indicated the dominance of the $\langle VVP \rangle$ structure in the production vertex, 
with a possible admixture of $\langle VV \rangle \langle P \rangle$. With the limited experimental information that we have at our 
disposal we could confirm this hypothesis and reproduced the ratio $R_1 =  \mathcal{B}(\chi_{c1} \rightarrow \eta f_2(1270))/\mathcal{B}(\chi_{c1} \rightarrow \eta^{\prime}f'_2(1525))$ 
which unfortunately has still large errors. We could then determine six more ratios which are predictions of the model, some of 
which are more stable under the changes within the experimental uncertainties of the $R_1$ ratio. We hope that in the near future some of these ratios are measured to test the predictions of the model and the underlying 
hypothesis that the resonances considered are indeed dynamically generated from the $VV$ interaction.

\vspace{-0.3cm}
\begin{acknowledgments}
J. M. D. and N. I. acknowledge the hospitality of the Guangxi Normal University, China, where this work was carried out.
The work of N. I. was partly supported by JSPS Overseas Research Fellowships and JSPS KAKENHI Grant Number JP19K14709.
This work is partly supported by the National Natural Science Foundation of China under Grants Nos. 11565007, 11847317, and 11975083.
This
work is also partly supported by the Spanish Ministerio
de Economia y Competitividad and European FEDER
funds under Contracts No. FIS2017-84038-C2-1-P B and
No. FIS2017-84038-C2-2-P B, and the project Severo
Ochoa of IFIC, SEV-2014-0398. 
\end{acknowledgments}

%\newpage

  \end{document}